%% file: ms.tex
\documentclass[conference]{IEEEtran}
\IEEEoverridecommandlockouts
\usepackage{cite}
\usepackage{amsmath,amssymb,amsfonts}
\usepackage{algorithmic}
\usepackage{graphicx}
\usepackage{textcomp}
\usepackage{xcolor}
\usepackage{float}

\def\BibTeX{{\rm B\kern-.05em{\sc i\kern-.025em b}\kern-.08em
    T\kern-.1667em\lower.7ex\hbox{E}\kern-.125emX}}
\begin{document}

\title{ConvNeXt-backbone HoVerNet for Nuclei Segmentation and Classification\\ {\large Using Haematoxylin Composition to Smooth One-Hot Label}}


\makeatletter
\newcommand{\linebreakand}{%
  \end{@IEEEauthorhalign}
  \hfill\mbox{}\par
  \mbox{}\hfill\begin{@IEEEauthorhalign}
}
\makeatother

\author{
  \IEEEauthorblockN{Jiachen Li, Chixin Wang, Banban Huang}
  \IEEEauthorblockA{\textit{School of software engineering} \\
    \textit{South China University of Technology}, Guangzhou, China\\
    {202021046291, 202121047134, 202021046278}@mail.scut.edu.cn}
  \and
  \IEEEauthorblockN{Zekun Zhou}
  \IEEEauthorblockA{\textit{School of Computer Science and Cyber Engineering} \\
    \textit{Guangzhou University}, Guangzhou, China \\
    2112006294@e.gzhu.edu.cn}
}

\maketitle

\begin{abstract}
This manuscript gives a brief description of the algorithm used to participate in CoNIC Challenge 2022 \cite{graham2021conic}. After the baseline was made available, we follow the method in it and replace the ResNet baseline with ConvNeXt one. Moreover, we propose to first convert RGB space to Haematoxylin-Eosin-DAB(HED) space, then use Haematoxylin composition of origin image to smooth semantic one hot label. Afterwards, nuclei distribution of train and valid set are explored to select the best fold split for training model for final test phase submission. Results on validation set shows that even with channel of each stage smaller in number, HoVerNet with ConvNeXt-tiny backbone still improves the mPQ+ by 0.04 and multi r2 by 0.0144.

\end{abstract}


\section{Introduction}
The past decades have witnessed substantial progress for deep neural network. With the help of deep learning, regions of interest can be easily obtained by automatic methods in segmentation tasks within seconds or minutes. Therefore, there emerges a great number of deep learning assisted method for medical image analysis. 

In computational pathology, nuclei segmentation is essential for cancerous tissue study, as well as downstream task such as cancer screening, cancer grading and caner type prediction. However, a whole slide image contains tens of thousand of nuclei. Inter- and intra instance of different types' nuclei vary not only in size and shape, but also in surroundings, scanning setup (devices and light condition) and Haematoxylin and Eosin (H\&E) stained effect. So It's irrealistic for practitioners to manually segment such numerous nuclei under this circumstance. 

\section{Related Work}

\subsection{Feature Map for Nuclei Segmentation}

By learning horizontal and vertical feature map and nuclei pixel map, HoVerNet\cite{graham2019hover} separate clustered nuclei instances using watershed algorithm, and jointly segment the nuclei type with a third feature map name nuclei types.
CDNet\cite{he2021cdnet} proposed centripetal direction feature and direction difference map to tackle overlapped nuclei.
Similarly, CPP-Net\cite{chen2021cpp} learned k-th direction pixel to boundary distance using context-aware polygon proposal.
CS-Net\cite{peng2021cs} introduced an instance-aware loss and used boundary attentive learning to identify every instances.
StarDist\cite{schmidt2018cell} proposed to use star-convex polygons to detect cells, predicting radial distances to their center.

\subsection{Model Structure for Nuclei Segmentation}
Most existing methods follow the paradigm of U-Net\cite{ronneberger2015u}, which first extract features from high to low resolution, and then restore the resolution by upsampling and concatenating. Aforementioned method mainly segment nuclei instances by post prossessing semantic feature. Different from that, Vuola \cite{vuola2019mask} uses Mask-RCNN to segment nuclei, in which nuclei are first detected and then segmented into several types.

\section{Method}
We first conducted experiments using Swin-Transformer for semantic segmentation. Sadly, it's not such good as baseline method as HoVerNet for it mainly extracting features in a more lower resolution (4 times smaller). When the resolution is modified the same as HoVerNet, attention mechanism can not bare its complexity as the computation are square of 4 (16) times larger.

Therefore, using attention based architecture as a backbone feature extractor is realistic. However, it's still feasible to embedded transformer somewhere later in or after ResNet extension, such as replacing the dense block sub-module. Owning to inefficient time, we left it for later research.

Another state of the art convolution architecture, named ConvNeXt\cite{liu2022convnet} has recently publish as a competitor to attention based model (refers to Transformer most of the time). It takes the advantages of Swin-Transformer architecture designing pattern. We replace the ResNet-50 backbone in HoVerNet with ConvNeXt-tiny, obtaining a slightly performance increase. We will talk about it later in method description.

\begin{figure*}
\centering   
\includegraphics[width=0.9\textwidth]{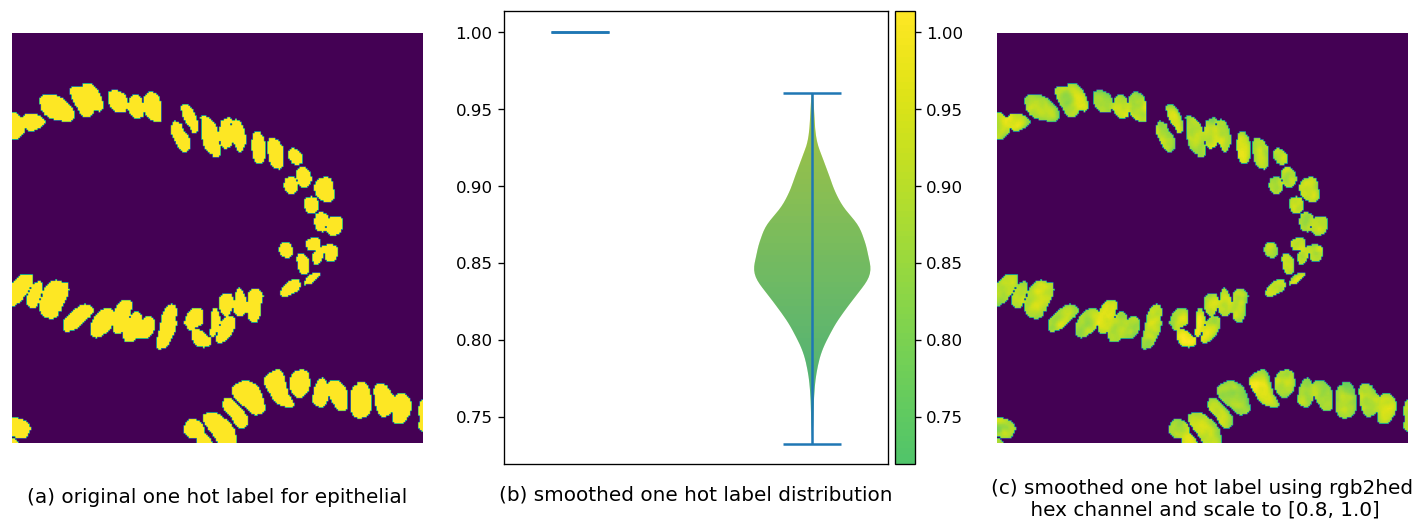}
\caption{One hot label smoothing using haematoxylin channel decomposition} 
\label{fig:1} 
\end{figure*}

\subsection{Data prepossessing and augmentation}\label{AA}
In the pipeline of hovernet, the part of instance segmentation is involved the distance between data points and cell centre. The instance feature map is prepossess into horizontal and vertical feature map and nuclei pixel map as mentioned in HoVerNet. The method of prepossessing mainly lies in the ground truth labels which help the algorithm identify nuclei instances. 

\textbf{Instance segmentation: HV feature map}. Taking the horizontal feature map as an example. The pixels are cropped from a nucleus instance. After that, center of nucleus mass is obtained. The left side of center of mass is normalized into [-1, 0] and the right side [0, 1], scaled by the distances to minimum and maximum of horizontal coordinates.

\textbf{Semantic segmentation: RGB to HED space conversion}. As \textbf{Fig.~\ref{fig:2}} reveals, \cite{ruifrok2001quantification} propose a way to decompose the RGB space H\&E stained images into Haematoxylin-Eosin-DAB(HED) space. The DAB composition neglected for simplicity. We first use Skimage.color.rgb2hed to perform the conversion. Then only the Haematoxylin composition(marked as $H$) is taken out for further operation because it indicates the spatial density of DNA stained. \textbf{One hot label smoothing} $H$ is rescale to [0,1] and then $0.5+0.5*H_{norm}$ to replace the 1s in original nuclei one hot label for each category \textbf{Fig.~\ref{fig:1}}. The smoothed one hot label can be written as follow:
\begin{equation*}
\begin{aligned}
    \textbf{HED} = rgb2hed(\textbf{IMG}[i,...]), \textbf{H} = \textbf{HED}[...,0] \\
    \textbf{H}_{norm} = Maxmin_{(0,1)}(\textbf{H}), \textbf{H}^{\prime} = 0.5 + 0.5*\textbf{H}_{norm} \\
    \textbf{L} = OneHot(\textbf{LBL}[i,...,1]), \textbf{L}^{\prime} = \textbf{L}*\textbf{H}^{\prime} \\
\end{aligned}
\end{equation*}



\subsection{Method description}
Our method still obeys the pipeline of HoVerNet, with some of components altered as an inspiration of new publication. The model \textbf{Fig.~\ref{fig:3}} we are using contains a ConvNeXt backbone for feature extraction, where multi-resolution features are output in specific stages, forming a feature pyramid. After that, three segmentation heads are embedded in a U-Net like structure to learn nucleus pixel, horizontal and vertical distances and nucleus type, which indicates whether a pixel belongs to a foreground nucleus, distances to the nucleus center mass and which type of nucleus it is respectively. The former two feature map are responsible for nucleus instances segmentation using watershed in post-processing phase. And the latter one is used as semantic segmentation straightly. It's worth notice that we use an ensemble of focal loss \cite{lin2017focal} and dice loss to train tp branch, where $loss_{tp} = 3*focal + dice$.

\begin{figure}
\centering   
\includegraphics[width=0.48\textwidth]{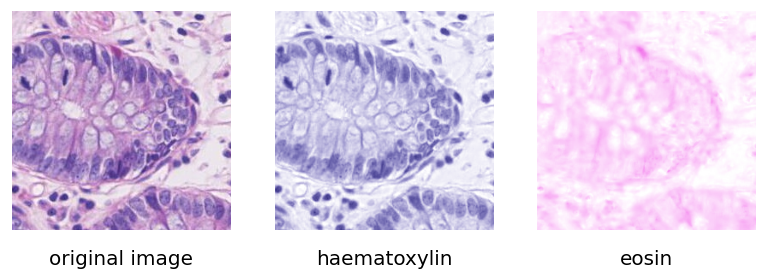}
 \caption{Haematoxylin and eosin (H\&E) decomposition of origin image} 
\label{fig:2} 
\end{figure}

\subsection{Post Processing}
The nucleus pixel feature map is first passing through a threshold , turning the pixel with a probability over 0.5 into 1. Small object under size of 10 are also removed in this phase. And we got a 0-1 feature map indicates whether a pixel belongs to a foreground nucleus. After that, the maximum between pixel horizontal and vertical is representative as the distance to nucleus center of mass. It also means how much height it is to the lowest point of nuclei basin, as center of nucleus mass remains 0. And the watershed algorithm is executed on these post processing result to produce the final instance segmentation result.

\begin{table}[htbp]
\caption{Result of model for FOLD 1 Validation set}
\begin{center}
\begin{tabular}{|c|c|c|c|}
\hline
\textbf{HoVerNet}&\multicolumn{3}{|c|}{\textbf{Indicators}} \\
\cline{2-4} 
backbone & \textbf{\textit{PQ}}& \textbf{\textit{mPQ+}}& \textbf{\textit{Multi-R2}} \\
\hline
ResNet-50-Ext & 0.6149 & 0.4998 & 0.8585 \\
\hline
ConvNeXt-Tiny-Ext & 0.6106 & 0.5040 & 0.8729 \\
\hline
\end{tabular}
\label{tab1}
\end{center}
\end{table}

\begin{figure*}[bhtp]
\centering   
\includegraphics[width=0.8\textwidth]{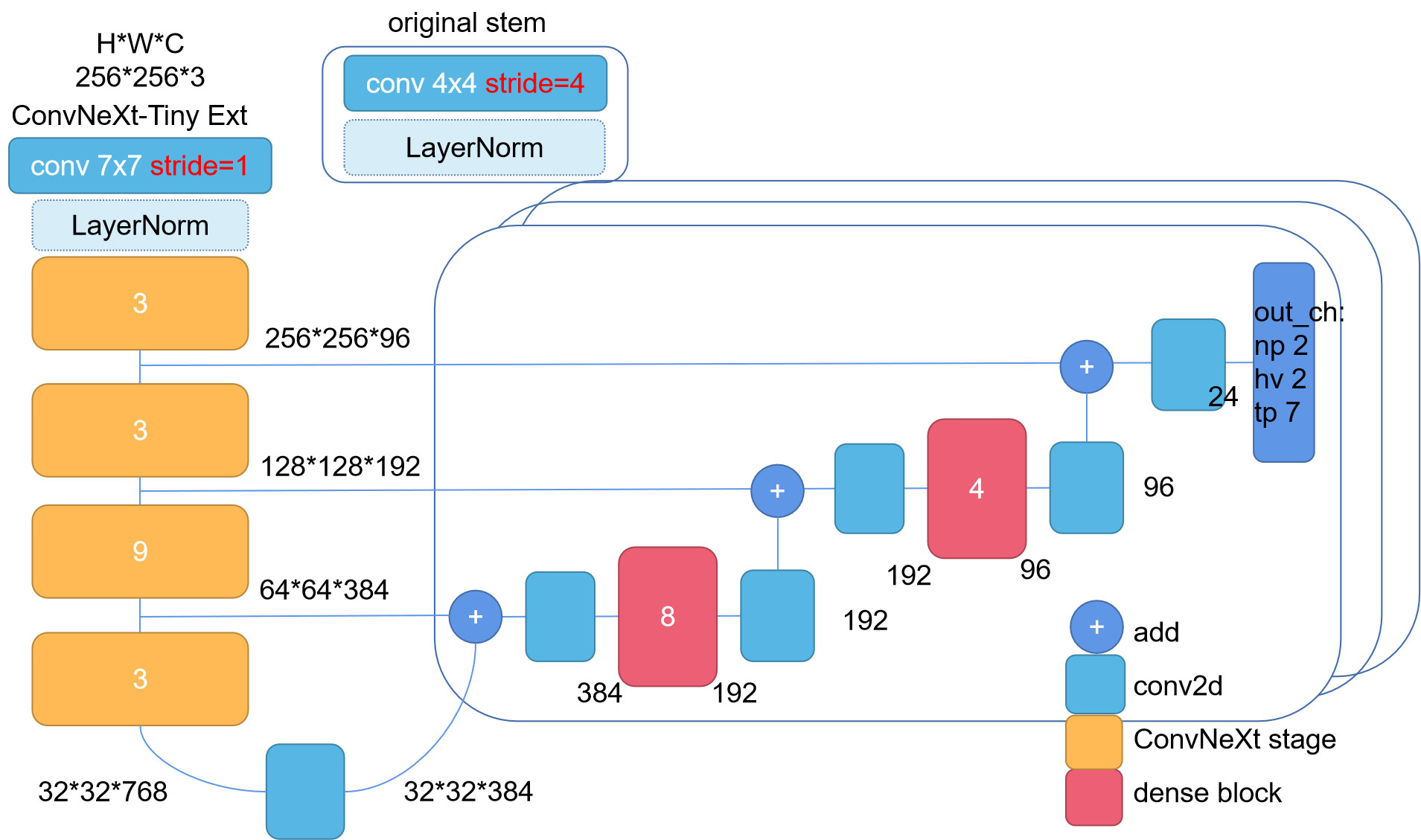}
\caption{Model structure used to participate in CoNIC Challenge 2022}
\label{fig:3}
\end{figure*}

\subsection{Results}
Experiments are conducted on 2 NVIDIA 2080Ti GPU, batch size is set to 4 for both training and validation. It took around 12 hours to run the whole pipeline. Results are shown in Tab.~\ref{tab1}. 

Out of inefficient time and energy, we were unable to perform 10 fold cross validation on Lizard dataset \cite{graham2021lizard} divided by the official CoNIC HoVerNet testing baseline. But we investigate the proportion of each category nuclei type variations. The number of each cell are counted and percentage are calculated. Then the train to valid ratio of each nuclei proportion type is obtained as shown in \textbf{Fig~.\ref{fig:4}}. 

For each category of nuclei $i$, we first sum up all its counts in train and valid set $C_{i, train}$, $C_{i, valid}$ respectively. Compute its ratio over total cell counts $R_{i, train}=C_{i, train}/\sum C_{i, train}$, and obtain train to valid distribution similarity using $Sim_{i} = R_{i, train} / R_{i, valid}$. The closer it's to 1, the more identical distribution of nuclei type $i$ in train and valid set is. Eventually, we select fold 06 based on its minimum $\sum (Sim_{i}-1)^{2}$ legend on \textbf{Fig~.\ref{fig:4}}.

\section*{Conclusion}
HoVerNet is used as baseline method. And the feature extraction backbone is altered with ConvNeXt, which brings slightly improvement to model performance. Experiments shows that even  with channel of each stage significant smaller in size, convnext backbone still improves model performance, meaning that there is still redundancy in channel dimension. Therefore, we will make the architecture channels searchable in the future. Use a neural architecture search method to jointly optimize performance and throughput, make it more feasible to deploy on edge devices.

\begin{figure}[htp]
\centering   
\includegraphics[width=0.48\textwidth,trim=15 0 0 5,clip]{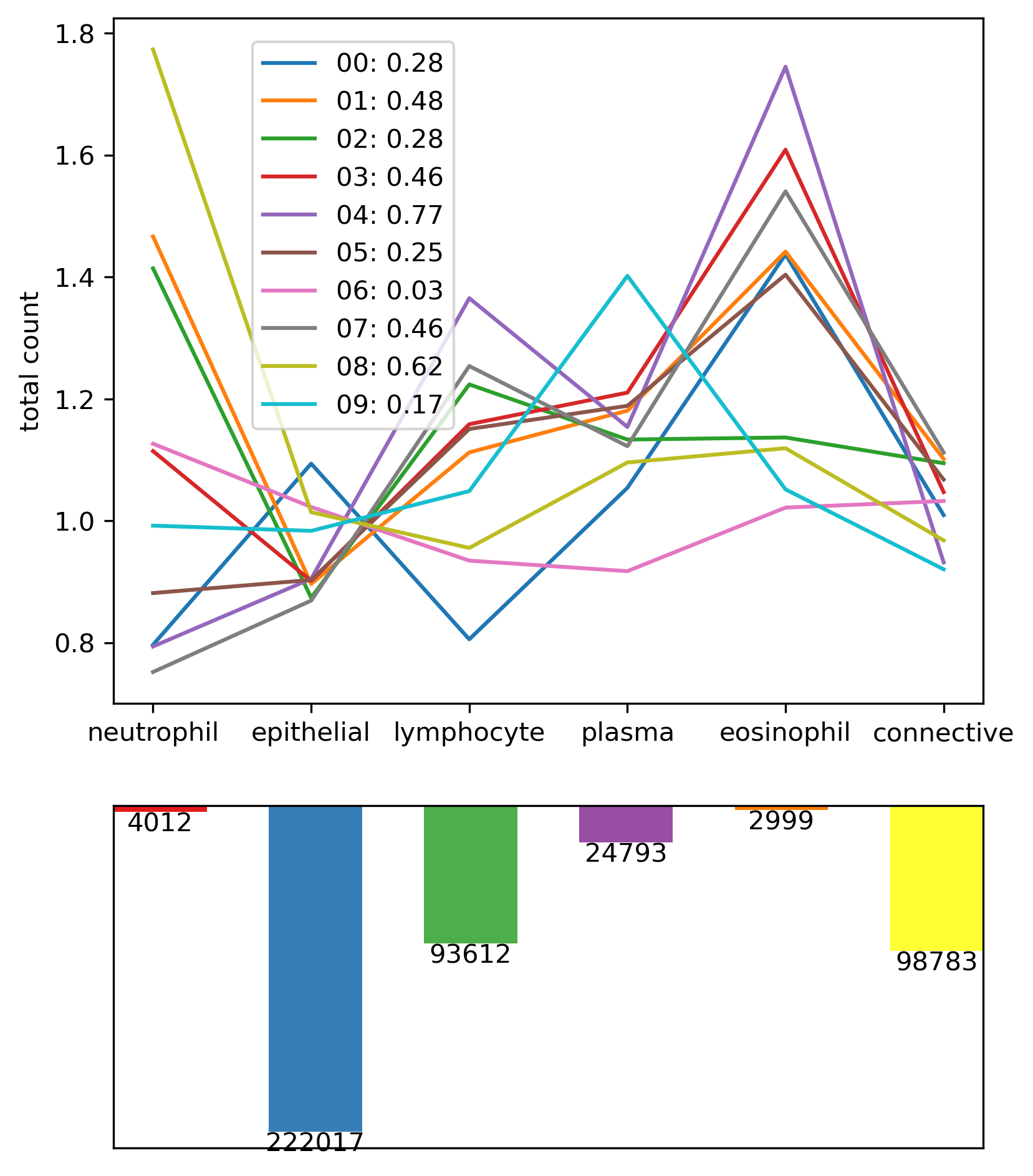}
\caption{Train valid fold cell count ratio and nuclei category distribution} 
\label{fig:4} 
\end{figure}


\input{ms.bbl}



\begin{figure*}
\centering   
\includegraphics[width=0.8\textwidth]{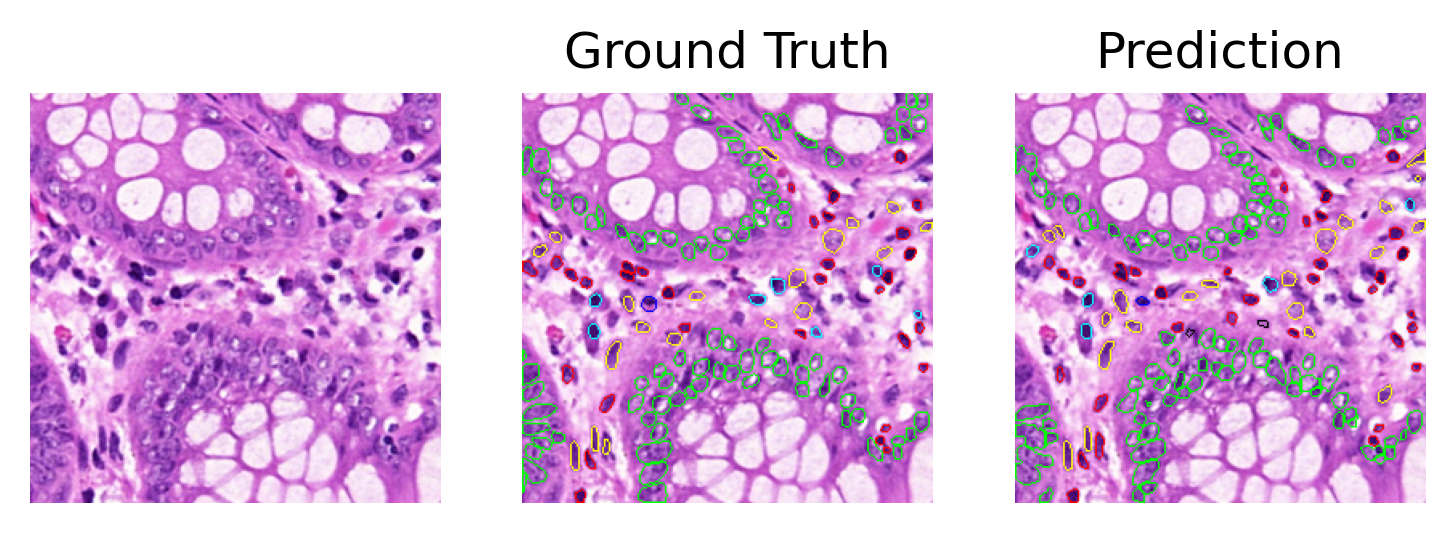}
\includegraphics[width=0.8\textwidth,trim=0 0 0 18,clip]{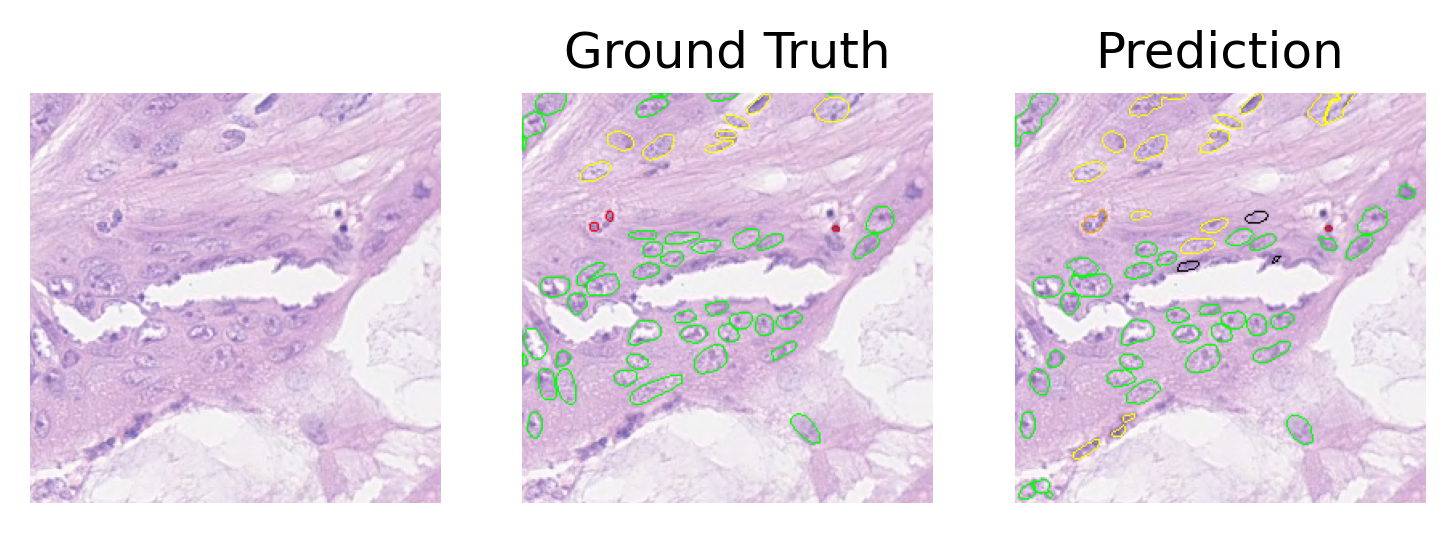}
\includegraphics[width=0.8\textwidth,trim=0 0 0 18,clip]{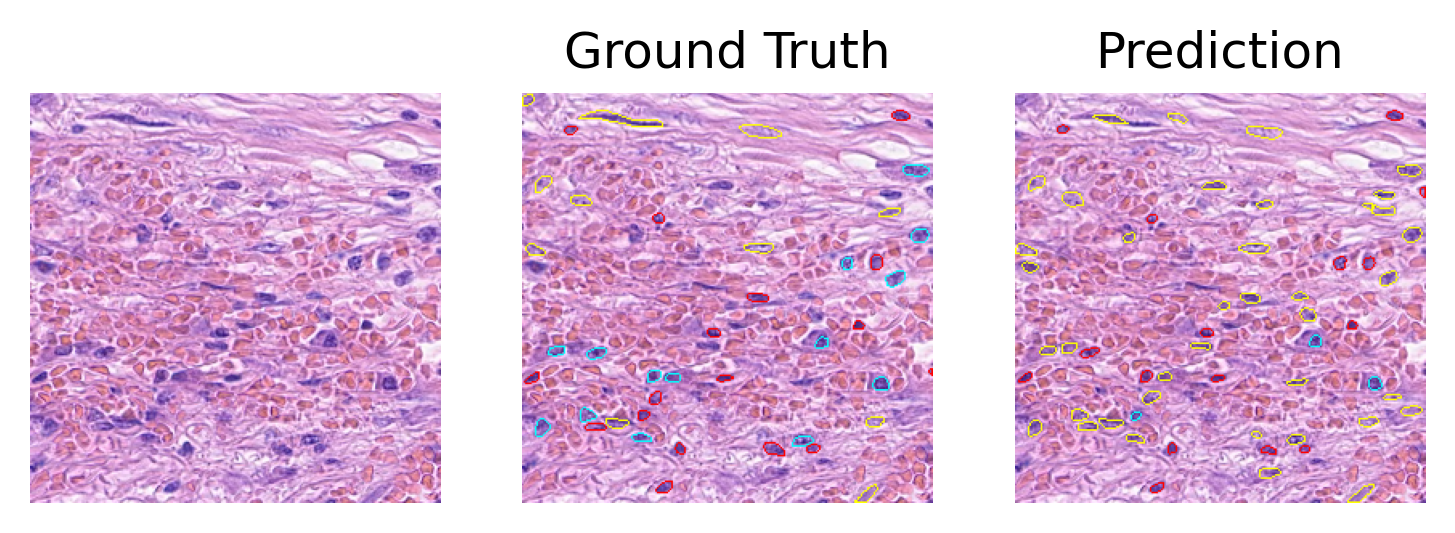}
\includegraphics[width=0.8\textwidth,trim=0 0 0 18,clip]{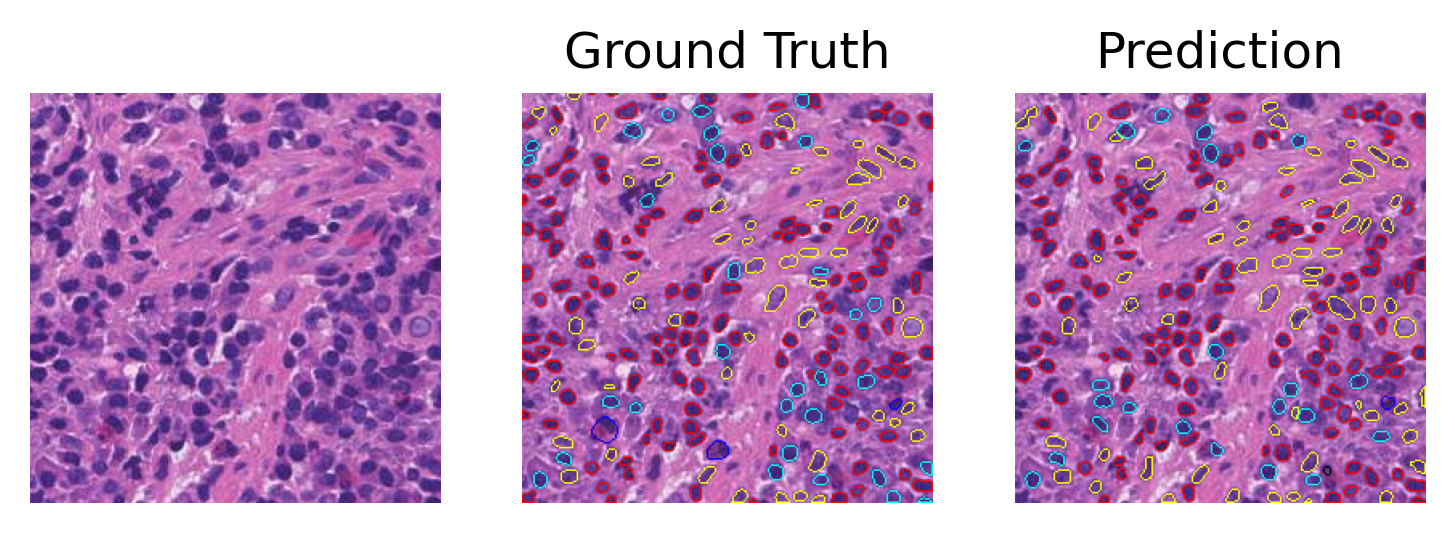}
\caption{Ground truth and model prediction on Fold 1 validation set}
\end{figure*}

\end{document}